\documentclass[11pt]{article}

\usepackage{authblk} % author list

\usepackage[T1]{fontenc}
\usepackage[utf8]{inputenc}
\usepackage{newtxtext}
\usepackage{newtxmath}
\usepackage{microtype}
\usepackage{booktabs}
\usepackage{placeins}
\usepackage{longtable}
\usepackage{array}
\usepackage{calc}
\usepackage{geometry}
\usepackage{todonotes}
\usepackage{graphicx}
\usepackage{hyperref}
\usepackage{caption}
\usepackage{tikz}
\usetikzlibrary{arrows, positioning, decorations.pathreplacing}
\usepackage{xcolor}
\usepackage{tablefootnote}
\usepackage{tabularx}
\usepackage[markup=default, final]{changes}
\usepackage{subcaption}
\definechangesauthor[color=orange]{AL}

\usepackage{natbib}
\usepackage[ttdefault=true]{AnonymousPro}
\usepackage{float}

\usepackage{setspace}
\makeatletter
\newif\if@anonymize
\@anonymizefalse

\newcommand{\blinded}[2]{%
\if@anonymize
#2%
\else
#1%
\fi}
\title{A Variance Decomposition Approach to Inconclusives in Forensic Black Box Studies}

\if@anonymize
\author{(Blinded authors)}
\else
\author[1]{Amanda Luby}
\author[2, *]{Joseph B Kadane}
\affil[1]{Department of Mathematics and Statistics, Carleton College}
\affil[2]{Department of Statistics and Data Science, Carnegie Mellon University} 
\affil[*]{Corresponding author: Joseph B Kadane, kadane@stat.cmu.edu}

\date{}
\fi

\begin{document}

\maketitle

\begin{abstract}
  In the US, `black box' studies are increasingly being used to estimate the error rate of forensic disciplines. A sample of forensic examiner participants are asked to evaluate a set of items whose source is known to the researchers but not to the participants. Participants are asked to make a source determination (typically an identification, exclusion, or some kind of inconclusive). We study inconclusives in two black box studies,  one on fingerprints and one on bullets. Rather than treating all inconclusive responses as functionally correct (as is the practice in reported error rates in the two studies we address), irrelevant to reported error rates (as some would do), or treating them all as potential errors (as others would do), we  propose that the overall pattern of inconclusives in a particular black box study can shed light on the proportion of inconclusives that are due to examiner variability. Raw item and examiner variances are computed, and compared with the results of a logistic regression model that takes account of which items were addressed by which examiner. The error rates reported in black box studies are substantially smaller than ``failure rate" analyses that take inconclusives into account. The magnitude of this difference is highly dependent on the particular study at hand.

  \noindent \textbf{Keywords:} Black Box Studies, Error Rates, Firearms, Fingerprints, Logistic Regression
\end{abstract}

\section{Introduction}\label{introduction}

Both the reports of \citet{nrc2009} and the
\citet{pcast2016}
highlighted the importance of `black box' studies to estimate the error
rate of subjective forensic methods. In these studies, forensic examiners function as `black boxes' who make source determinations without revealing their underlying reasoning. Accuracy is measured by comparing their conclusions to the known ground truth (same source or different source).

Typically, examiners can make an identification (same source
conclusion), an exclusion (different source conclusion), or an
inconclusive (neither an identification nor an exclusion). While some disciplines may use reporting scales with more than three categories, they can typically still be grouped into these three broad categories. Numerous `black box' studies have been conducted 
across forensic disciplines including latent prints \citep{ulery2011, ELDRIDGE2021110457}, firearms \citep{mattijssen2020validity, monson2022accuracy}, handwriting \citep{hicklin2022accuracy}, and
bloodstain pattern analysis \citep{hicklin2021accuracy}.

The first publication of such studies often estimate the error rate by dividing the number of erroneous determinations by the total number of determinations, including inconclusive results. A debate
has emerged regarding how inconclusive results should be treated \citep{dror2019cannot, hofmann2020treatment}. Some argue that inconclusives are potential errors \citep{dror2020mis, dror2022, dorfman2022inconclusives}, while others argue they may be ignored for computing error rates \citep{arkes-koehler2021, arkes-koehler2022}. The debate is heightened by the observation \citep{scurich2022, sinha-gutierrez2023} of high rates of inconclusives, particularly
in firearm test results. However, these papers largely
consider a single inconclusive at a time, and assume all inconclusive
responses should be treated identically as functionally correct, potentially erroneous, or ignorable. An inconclusive result is ambiguous: does it say something about the forensic examiner (and hence is possibly an error) or does it say something about the test item offered to the forensic examiner (and hence arguably benign)? We write with the premise that an inconclusive determination arises from the interaction between the examiner and the item, and hence reflects both. Analyzing the pattern of inconclusive responses across examiners and items can inform the debate. \added{Our focus is on the interpretation of large-scale black box study results, rather than an inconclusive arising in casework.}

By analyzing the variance in inconclusive rates \added{within a particular black box study}, we can estimate the proportion of inconclusives attributable to examiners variability and the proportion attributable to item characteristics. This approach avoids the pitfalls of treating all inconclusives uniformly, either as correct, incorrect, or ignorable. 

\begin{figure}

{\centering \includegraphics[width=\linewidth]{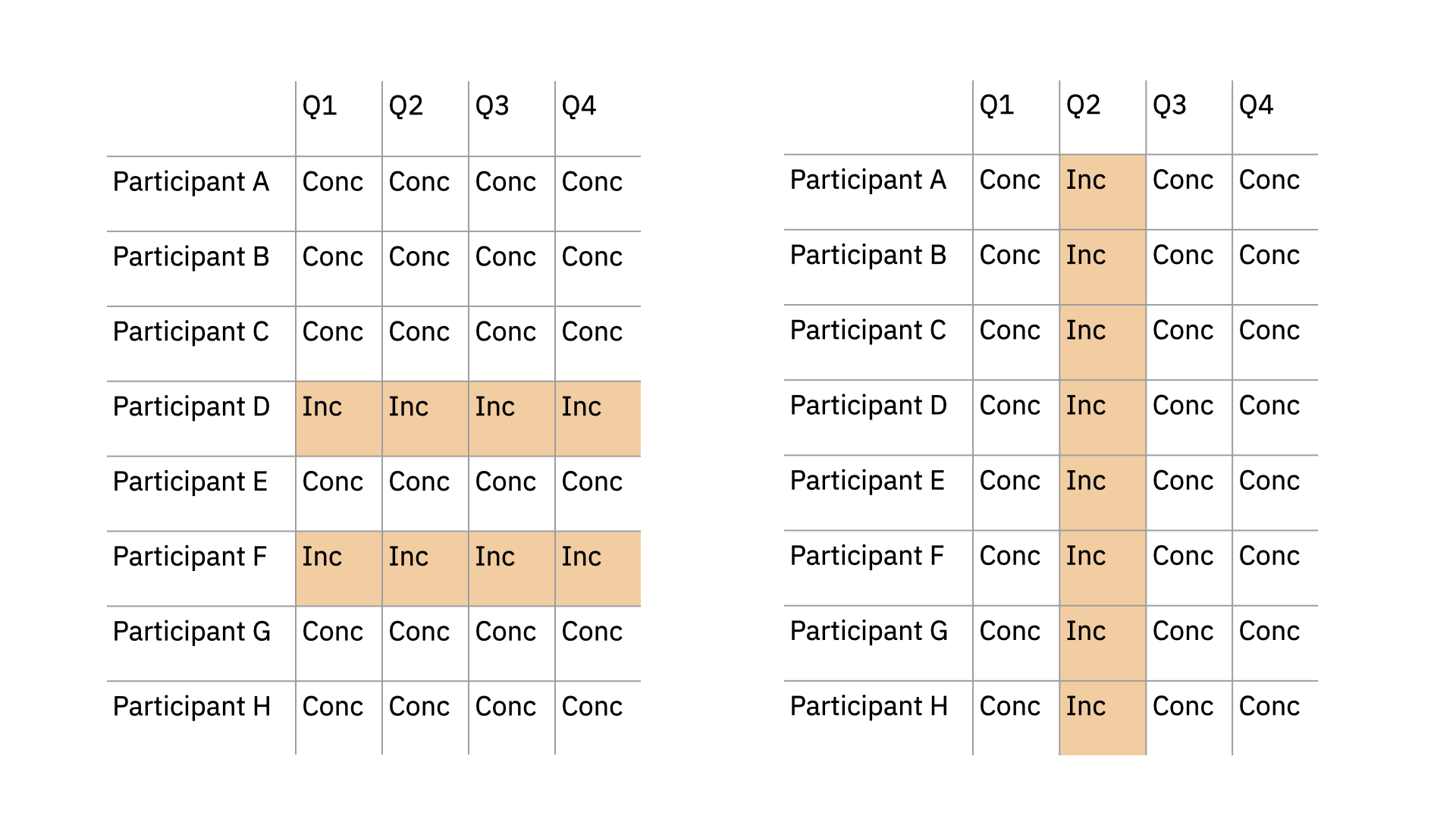}

}

\caption{\label{fig-hyp-study} Hypothetical results from Case I (left) and Case II (right). }

\end{figure}

By contrast, imagine two studies, in each of which participants decided that 25\% of the test items
were neither identification nor exclusions. In Case I, two of the
eight participants reported only inconclusives while all other participants reported only conclusive results. In Case II, one of the four items
was rated as inconclusive by every participant, while all other items were rated as conclusive by every participant. These hypothetical results are shown in Figure \ref{fig-hyp-study}. In Case I, the
participant inconclusive variance is large while the item inconclusive variance is zero, and it is
reasonable to regard inconclusives as a matter of the participant. In Case II, the participant inconclusive variance is zero while the item inconclusive variance is large, and the inconclusives can be regarded as due to the items chosen for the study. Of course, real world data does not follow either extreme.

We propose to attribute to the participants the fraction of
inconclusives equal to the ratio of the examiner variance to the total variance (i.e.~the sum of examiner variance and item variance). This ratio is equal to zero when there is perfect examiner agreement (e.g. Case II above) and equal to one when there is perfect item agreement (e.g. Case I above). The rationale is to separate decisions of the study designers (high item inconclusive variance) from determinations by some participants (high participant inconclusive variance). 

We do not treat every inconclusive identically as correct, incorrect, or ignorable (as in \citet{hofmann2020treatment}); nor do we score each inconclusive individually (as in the `wisdom of the crowds' approach of \citet{dror2020mis} or the `answer key' approach of \citet{luby2020behaviormetrika}). Rather, we use the overall patterns of inconclusives in each particular study to weight some proportion of the inconclusives as attributable to examiners (and therefore potential errors) and some proportion as attributable to the items (and therefore attributable to the study designers). This approach provides context for the level of agreement among participants in a given study. Since different studies will naturally result in different levels of agreement, this framework provides a mechanism for comparing study results within and across disciplines. 

However, \added{this method relies on some amount of concentration of inconclusives on certain items, and may produce counterintuitive results in some edge cases.} Furthermore, in many study designs, not every participant answers every question, and
so the fraction of inconclusives for each participant depends on the subset of items that they were shown. To address these issues requires a more refined model. In particular, it requires a parameter for each test item and a parameter for each
participant. By adapting models used in the analysis of standardized
testing, we estimate the tendencies of each participant to choose
``inconclusive,'' and how likely a given test item is to be rated as
inconclusive by the examiners.

The remainder of this paper is organized as follows:
Section~\ref{sec-two-black-box-studies} provides an overview of the
datasets provided by \citet{monson2022accuracy} and \citet{ulery2011}; 
Section~\ref{sec-inc-error-rates} displays different calculations for
error rates currently in the literature;
Section~\ref{sec-empirical-variance} introduces variance decomposition
as a framework for analyzing inconclusive responses and
Section~\ref{sec-model-based-variance} does so using a formal
statistical model; and Section~\ref{sec-discussion} discusses these
results in the broader context of forensic science.

\section{Two black box studies}\label{sec-two-black-box-studies}

We use results from
black box studies in two different forensic disciplines to illustrate this approach. From latent
prints, we use the \citet{ulery2011} study, and from firearms we use
the \citet{monson2022accuracy} study. \citet{monson2022accuracy} included test items
from \added{both} cartridge cases and from bullets; we focus on the results from
bullets.

These two studies are similar in some ways. They both constructed a
large item bank (744 and 228 items, respectively) of known ground-truth
\replaced{comparisons}{samples} and assigned participants a subset of the item bank. Both
studies claimed to include relatively difficult comparisons.
Participation was generally voluntary and the number of participants was
similar across studies (169 in latent prints; 173 in bullets).

However, there are a few notable differences. First, the item bank in
the latent prints study consisted of 70\% same-source items, while the
item bank in the bullets study contained only 17\% same-source items. In the latent prints study, low-quality latents were intentionally included to represent the range of items seen in casework. Reference prints for non-mated pairs were found using an IAFIS search, resulting in more `close non-matches' than in many other studies. The bullets study included items from three types of firearms and a single brand of ammunition. 30-60 `break-in' firings were used before generating test items in order to achieve `consistent and reproducible toolmarks'. 

In the latent prints study, each participant was assigned 98-110 items
(mode of 100); in the bullets study participants responded to 15 items
(\(n=59\)), 30 items (\(n=113\)), or 45 items (\(n=1\)). 68\% of all responses on the latent prints study were on same-source items; while 33\% of all responses on the bullets study were on same-source items. All practicing
latent print examiners were allowed to participate in the latent print
study, while the bullets study restricted participants to the United
States and excluded participants who worked for the FBI.

There are also subtle differences in the way conclusions were recorded
in each study, although source conclusions could generally be grouped
into \emph{same source} conclusions, \emph{different source}
conclusions, \emph{inconclusives}, and \emph{unsuitable} for
examination. The differences are outlined in
Table~\ref{tbl-conclusion-scales}.

If a response was inconclusive, both studies asked for further details.
The latent prints study asked participants to provide a reason for the
inconclusive, and the bullets study asked participants to report a type
of inconclusive on the AFTE scale \citep{afte1992theory}. The additional level of response can be grouped into
\emph{support for same source}, \emph{support for difference source}, or
\emph{support for neither} (see Table~\ref{tbl-inconclusive-scales}).

A more detailed comparison of the two studies is given in Appendix A.

\pagebreak

\begin{longtable}[t]{@{}
  >{\raggedright\arraybackslash}p{(\columnwidth - 4\tabcolsep) * \real{0.2000}}
  >{\raggedright\arraybackslash}p{(\columnwidth - 4\tabcolsep) * \real{0.4000}}
  >{\raggedright\arraybackslash}p{(\columnwidth - 4\tabcolsep) * \real{0.4000}}@{}}
\caption{\label{tbl-conclusion-scales}Conclusion scale for the two
studies.}\tabularnewline
\toprule\noalign{}
\begin{minipage}[b]{\linewidth}\raggedright
Conclusion
\end{minipage} & \begin{minipage}[b]{\linewidth}\raggedright
Ulery et al. (2011)
\end{minipage} & \begin{minipage}[b]{\linewidth}\raggedright
Monson et al. (2022)
\end{minipage} \\
\midrule\noalign{}
\endfirsthead
\toprule\noalign{}
\begin{minipage}[b]{\linewidth}\raggedright
Conclusion
\end{minipage} & \begin{minipage}[b]{\linewidth}\raggedright
Ulery et al. (2011)
\end{minipage} & \begin{minipage}[b]{\linewidth}\raggedright
Monson et al. (2022)
\end{minipage} \\
\midrule\noalign{}
\endhead
\bottomrule\noalign{}
\endlastfoot
Same Source (called \emph{Identification} in this paper) &
\textbf{Individualization}: \emph{The two fingerprints originated from
the same finger.} & \textbf{Identification}: \emph{Agreement of a
combination of individual characteristics and all discernible class
characteristics where the extent of the agreement exceeds that which can
occur in the comparison of toolmarks made by different tools and is
consistent with the agreement demonstrated by toolmarks known to have
been produced by the same tool.} \\
Different Source (\emph{Exclusion}) & \textbf{Exclusion:} \emph{The two
fingerprints did not come from the same finger.} & \textbf{Elimination}:
\emph{The significant disagreement of discernible class characteristics
and/or individual characteristics.} \\
Inconclusive & \textbf{Inconclusive}: \emph{Neither individualization
nor exclusion is possible.} & (see Table \ref{tbl-inconclusive-scales}) \\
Unsuitable & \textbf{No Value:} latent print image was not ``of value
for individualization'' or ``of value for exclusion only'' &
\textbf{Unsuitable} for examination \\
\end{longtable}

\pagebreak 

\hypertarget{tbl-inconclusive-scales}{}
\begin{longtable}[t]{@{}
  >{\raggedright\arraybackslash}p{(\columnwidth - 4\tabcolsep) * \real{0.2000}}
  >{\raggedright\arraybackslash}p{(\columnwidth - 4\tabcolsep) * \real{0.4000}}
  >{\raggedright\arraybackslash}p{(\columnwidth - 4\tabcolsep) * \real{0.4000}}@{}}
\caption{\label{tbl-inconclusive-scales}Inconclusive options for the two
studies}\tabularnewline
\toprule\noalign{}
\endfirsthead
\endhead
\bottomrule\noalign{}
\endlastfoot
& Ulery et al. (2011) & Monson et al. (2022) \\
Support for same source & \textbf{Close}: \emph{The correspondence of
features is supportive of the conclusion that the two impressions
originated from the same source, but not to the extent sufficient for
individualization.} & \textbf{Inc-A}: \emph{Some agreement of individual
characteristics and all discernible class characteristics, but
insufficient for an identification.} \\
Support for different source & \emph{NA} & \textbf{Inc-C}:
\emph{Agreement of all discernible class characteristics and
disagreement of individual characteristics, but insufficient for an
elimination.} \\
Support for neither same source nor different source &
\textbf{Insufficient}: \emph{Potentially corresponding areas are
present, but there is insufficient information present.} Examiners were
also told to select this reason if the reference print was not of value. 

\textbf{No Overlap}: \emph{No overlapping area between the latent and reference prints}
& \textbf{Inc-B:} \emph{Agreement of all discernible class
characteristics without agreement or disagreement of individual
characteristics due to an absence, insufficiency, or lack of
reproducibility.} \\
\end{longtable}

\pagebreak

For the purposes of this paper, we consider all non-conclusive responses
as inconclusives. In the latent prints study, participants could deem a
latent print to be of \emph{no value} before seeing the reference print,
or deem a pair of prints to be \emph{inconclusive} after seeing the
reference print. Similarly, we consider a response of \emph{unsuitable}
in the bullets study to also be \emph{inconclusive}. While these options
could have different operational impacts in casework and could represent
differing decision thresholds, we pool them together in this analysis,
since we are primarily interested in the overall tendencies to reach
conclusive decisions. 

In both studies, non-conclusive responses occur more often than
conclusive responses. A non-conclusive response is also predictive of the
ground truth of the item: for the latent prints data, the empirical
P(Same Source \textbar{} Not Conclusive) = 0.82. (See
Table~\ref{tbl-ulery-conc} (a)), and for the bullets data, P(Same Source
\textbar Not Conclusive) = 0.14 (See
Table~\ref{tbl-ulery-conc} (b)). After accounting for the differing base rates between same source and different source items in each study, the same trend holds. Inconclusive responses are 2.2 times more likely on same source items in the latent prints study, and only 0.33 times as likely in the bullets study. 
These trends could also be due to fundamental differences across disciplines, differences in item bank construction between the two studies (for example, the
bullets study consisted of firings from a single brand of ammunition), or
differences in the underlying behaviors of each sample of examiners.

\begin{table}

\caption{\label{tbl-ulery-conc}Frequency of conclusive vs not conclusive
responses}\begin{minipage}[t]{0.50\linewidth}
\subcaption{\label{tbl-ulery-conc-1}Latent Prints}

{\centering 

\begin{tabular}[t]{lrrr}
\toprule
 & Concl & Not Concl & Sum\\
\midrule
SS & 4314 & 7264 & 11578\\
DS & 3953 & 1590 & 5543\\
Sum & 8267 & 8854 & 17121\\
\bottomrule
\end{tabular}

}

\end{minipage}%
\begin{minipage}[t]{0.50\linewidth}
\subcaption{\label{tbl-ulery-conc-2}Bullets}

{\centering 

\begin{tabular}[t]{lrrr}
\toprule
 & Concl & Not Concl & Sum\\
\midrule
SS & 1117 & 312 & 1429\\
DS & 981 & 1910 & 2891\\
Sum & 2098 & 2222 & 4320\\
\bottomrule
\end{tabular}

}

\end{minipage}%

\end{table}

\begin{figure}

{\centering \includegraphics[width=\linewidth]{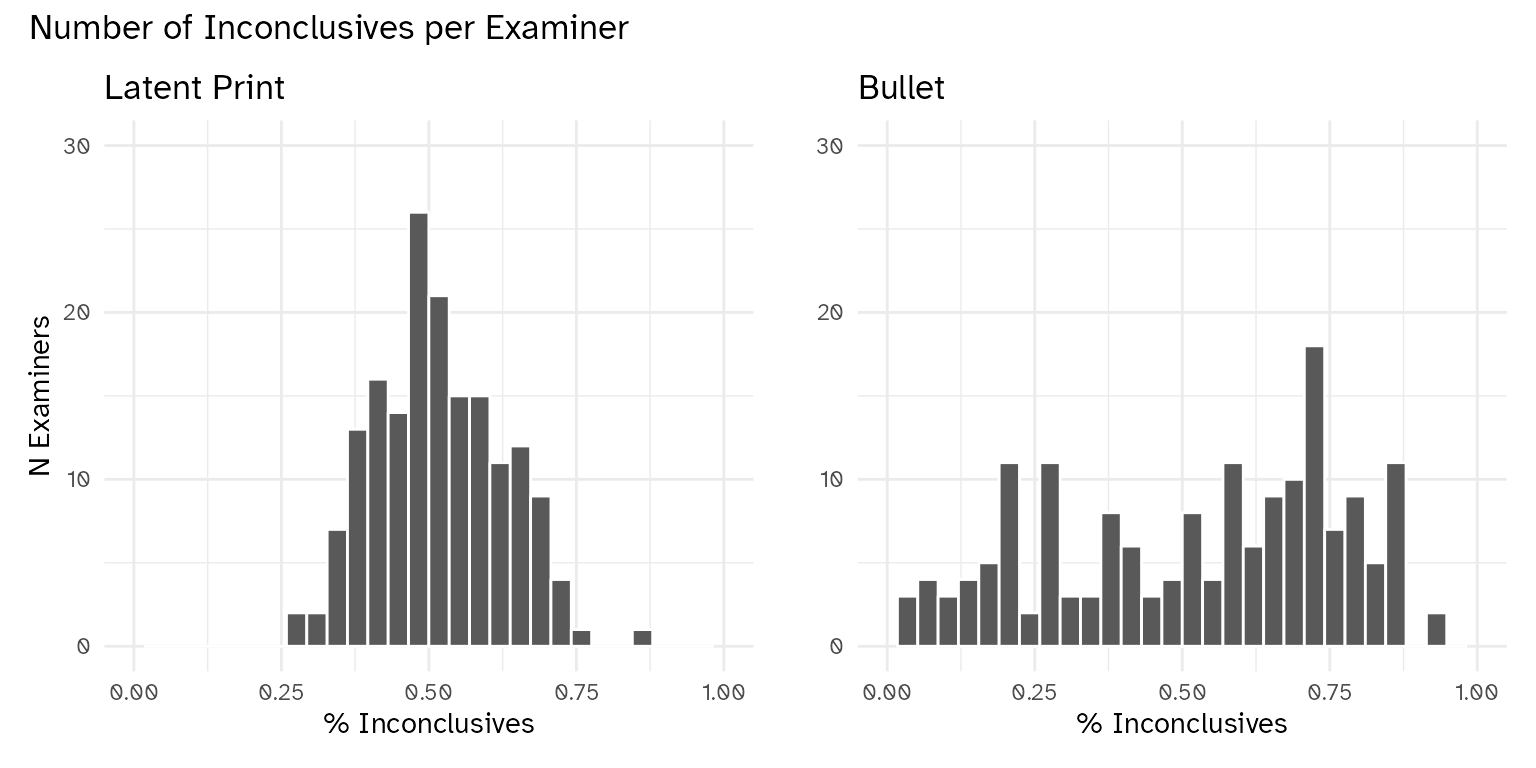}

}

\caption{\label{fig-incs-by-examiner}All examiners in the latent print
study reported at least 25\% inconclusives, with a mode of about 50\%.
Examiners in the bullets study were more varied, with some examiners
reporting 0 inconclusives and some reporting more than 90\%.}

\end{figure}

\begin{figure}

{\centering \includegraphics[width=\linewidth]{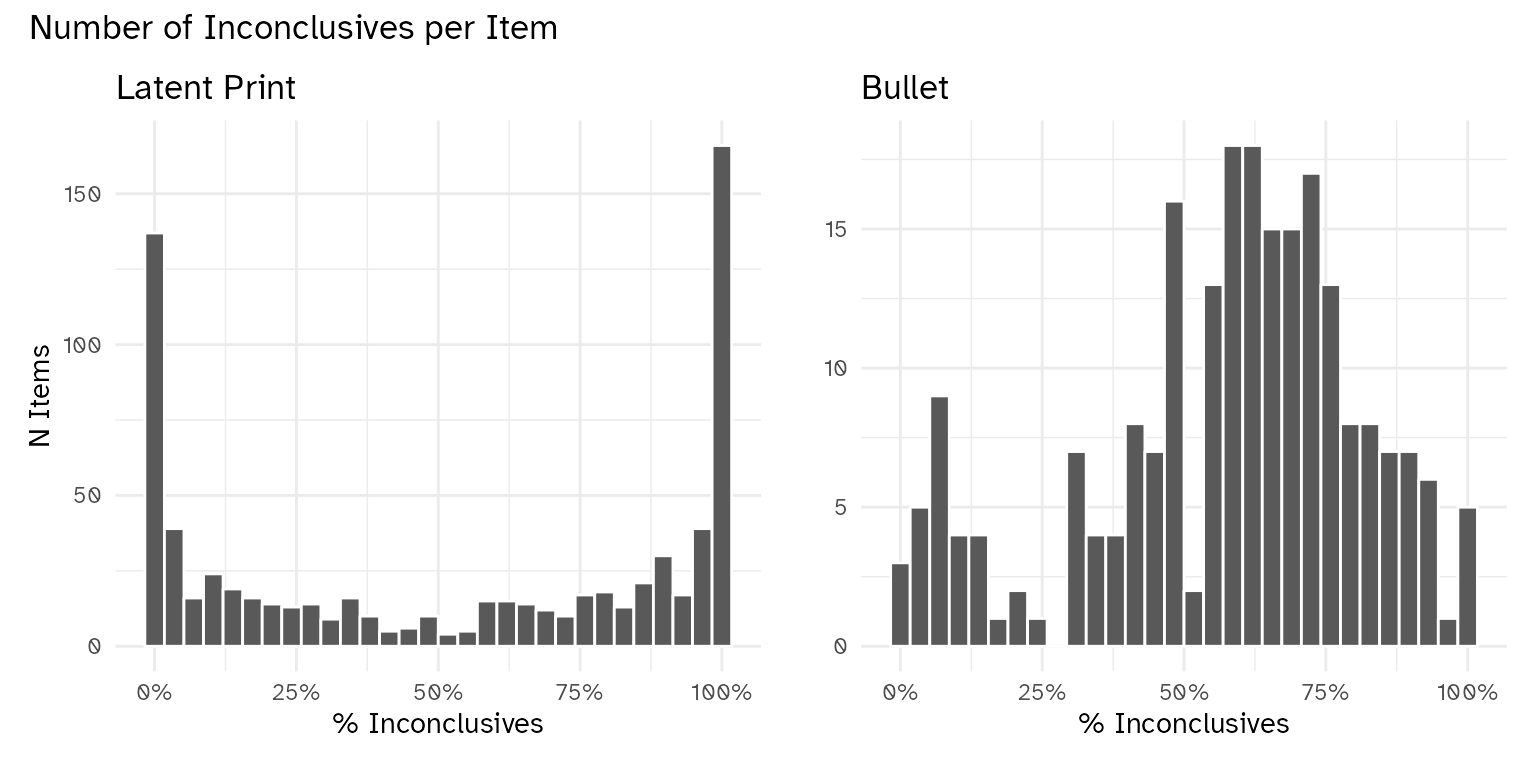}

}

\caption{\label{fig-incs-by-item}There are a large number of items in
the latent print study for which 0\% or 100\% of examiners who saw that
item reported it to be inconclusive, and other items are spread fairly
uniformly across the range of percentages. In the bullet study, most
items were reported to be inconclusive between 50\% and 75\% of the
time, and fewer items were regarded unanimously.}

\end{figure}

In the latent print study, examiners tend to concentrate around 50\%
inconclusive responses and range from 25\% to 85\% inconclusive. In the
bullets study, examiners were more varied, with some examiners reporting
zero inconclusives and some reporting more than 90\% inconclusive (see
Figure~\ref{fig-incs-by-examiner}). Many items in the latent prints
study were unanimously rated as conclusive or inconclusive, although items do
cover the entire range of inconclusive percentage. The bullets study
contained fewer items rated unanimously than the latent print study, and most
items were reported inconclusive between 25\% and 75\% of the time (see
Figure~\ref{fig-incs-by-item}).

\begin{figure}

{\centering \includegraphics[width=\linewidth]{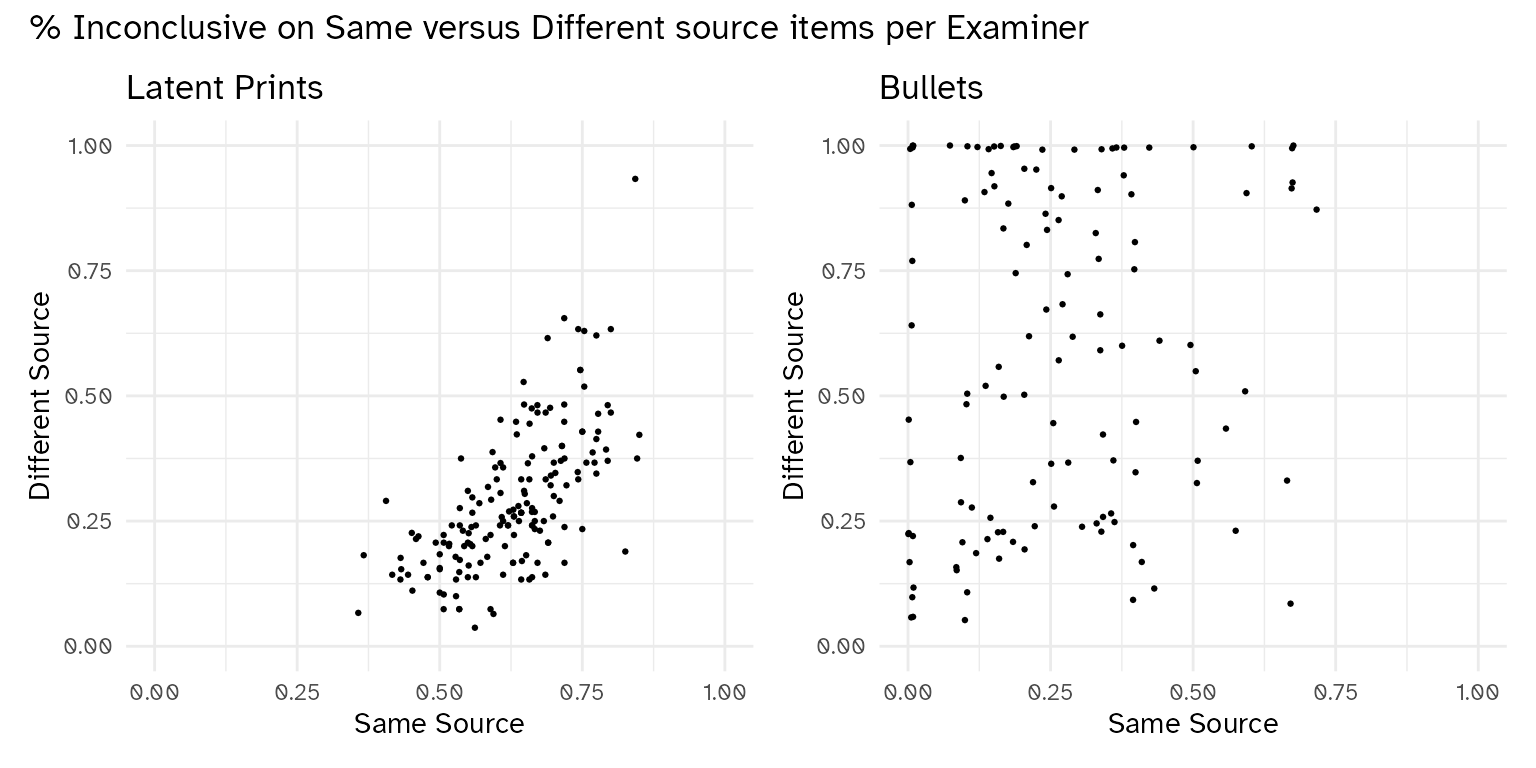}

}

\caption{\label{fig-incs-rates-by-examiner}In the latent print study,
examiners' tendency to be inconclusive on same source items was
correlated with their tendency to be inconclusive on different source
prints. In the bullets study, a number of examiners were never
inconclusive on same-source items and a number were always inconclusive
on different-source items.}

\end{figure}

Since we are primarily interested in the overall tendency of examiners
and items to be rated as inconclusive, rather than correctness, we do not
distinguish between correct conclusive decisions and errors. However,
there are some notable differences in behavior in the two studies on
different source versus same source items.
Figure~\ref{fig-incs-rates-by-examiner} shows that participants in the
latent print study who were likely to be inconclusive on same source
items were also likely to be inconclusive on different source items
(although inconclusives on same-source prints were overall observed more
frequently). Many participants in the bullets study, on the other hand,
displayed different behavior depending on whether the item was a same
source or different source item. There were a number of firearms
examiners who were never inconclusive on same source items, as well as a
number of examiners who were inconclusive on every different source item
that they saw.

While this could be due in part to the different subsets of items that
examiners were shown, it could also be due to very different thresholds
for inconclusive decisions on same versus different source prints. In
fact, the FBI has a policy that does not allow exclusions based on individual characteristics without access
to a physical firearm of the same make and model \citep{hofmann2020treatment}, and this may be the policy of other laboratories as well. This means that black box
studies in which only test fires and questioned bullets are provided to examiners will never result
in true exclusions OR false exclusions for those examiners. In the study
under consideration, FBI examiners were excluded from participating, but
if other laboratories also follow this policy, those examiners would not
necessarily have been excluded from participating. For this reason, we
perform our analyses on same-source and different-source comparisons
separately. When analyzing the different-source bullets data, we also
group examiners based on whether they (a) made \emph{any} eliminations
based on individual characteristics (across both bullets and cartridge
cases), or (b) did not make any eliminations based on individual
characteristics. 

\section{Impact of Inconclusives on Error
Rates}\label{sec-inc-error-rates}

While identifications and exclusions clearly can be labelled as correct
decisions or errors, it is unclear how inconclusive responses should enter the error rate. \citet{hofmann2020treatment} proposed Table~\ref{tbl-hofmann} to
summarize results from forensic black box studies and define error
rates. 

\begin{table}[]
\caption{\label{tbl-hofmann} \citet{hofmann2020treatment} summary of experimental
data.}\tabularnewline
\centering
\begin{tabular}{@{}llll@{}}
\toprule
 & Identification & Inconclusive & Exclusion \\ \midrule
Same Source & a & b & c \\
Different Source & d & e & f \\ \bottomrule
\end{tabular}
\end{table}

Below, we outline four possible error rate formulas. Option 1 and 2 are defined in \citet{hofmann2020treatment}, where the
difference lies in whether or not the inconclusive responses are
included in the denominator of the error rate. Option 3 treats
inconclusive responses as neither correct nor incorrect, but instead
assigns half credit \citep{luby2019decision}. Option 4 treats inconclusive responses as incorrect. FDA guidance suggests a procedure for handling inconclusive results in diagnostic tests \citep{FDA2007statistical}, which effectively amounts to using the "inconclusives correct" and "inconclusives incorrect" as lower and upper bounds on the error rate, respectively \citep{cuellar2024methodological}. The "half-credit" approach is presented as a possible point estimate between these two extremes.  
%Recall from Table~\ref{tbl-hofmann} that $a, b,$ and $c$ refer respectively to the number of Identifications, Inconclusives, and Exclusions among same source items; $d, e, f$ refer to the same among different source items.

\begin{enumerate}
\def\labelenumi{\arabic{enumi}.}
\item
  Inconclusives Ignored

  \begin{itemize}
  \item
    False Positive Error Rate = \(\frac{d}{d+f}\)
  \item
    False Negative Error Rate = \(\frac{c}{a + c}\)
  \end{itemize}
\item
  Inconclusives Correct

  \begin{itemize}
  \item
    False Positive Error Rate= \(\frac{d}{d+e+f}\)
  \item
    False Negative Error Rate = \(\frac{c}{a+b+c}\)
  \end{itemize}
\item
  Half Credit

  \begin{itemize}
  \item
    False Positive Error Rate = \(\frac{d + 0.5e}{d+e+f}\)
  \item
    False Negative Error Rate = \(\frac{c + 0.5 b}{a + b + c}\)
  \end{itemize}

\item Inconclusives incorrect
  \begin{itemize}
  \item
    False Positive Error Rate = \(\frac{d + e}{d+e+f}\)
  \item
    False Negative Error Rate = \(\frac{c + b}{a + b + c}\)
  \end{itemize}

\end{enumerate}

\begin{longtable}[t]{@{}lrrrr@{}}
\caption{\label{tbl-ulery-rates}Original and adjusted error rates for
the Ulery et al. (2011) latent print data. 'No Value' determinations have been excluded from these calculations, following the original study.}\tabularnewline
\toprule\noalign{}
Ground Truth & Inconc Ignored & Inconc Correct & Inconc Incorrect & Half Credit \\
\midrule\noalign{}
\endfirsthead
\toprule\noalign{}
Ground Truth & Inconc Ignored & Inconc Correct & Inconc Incorrect & Half Credit \\
\midrule\noalign{}
\endhead
\bottomrule\noalign{}
\endlastfoot
Different Source & 0.002 & 0.001 & 0.21 & 0.14 \\
Same Source & 0.142 & 0.075 & 0.55 & 0.37 \\
\end{longtable}

\FloatBarrier

\begin{table}[]
\centering
\caption{\label{tbl-monson-rates}Original and adjusted error rates for
the Monson et al. (2022) bullet data. \added{'Unsuitable' determinations have been excluded from these calculations, following the original study.} Note that examiners have been
grouped based on whether they made any eliminations based on individual
characteristics among the bullet and cartridge case sets they were
assigned.}
{%
\begin{tabular}{@{}llllll@{}}
\toprule
Ground Truth & Group & \begin{tabular}[c]{@{}l@{}}Inconc  Ignored\end{tabular} & \begin{tabular}[c]{@{}l@{}}Inconc  Correct\end{tabular} & \begin{tabular}[c]{@{}l@{}}Inconc  Incorrect\end{tabular} & \begin{tabular}[c]{@{}l@{}}Half  Credit\end{tabular} \\ \midrule
\begin{tabular}[c]{@{}l@{}}Different \\ Source\end{tabular} & \begin{tabular}[c]{@{}l@{}}Made \\ Eliminations\end{tabular} & 0.018 & 0.008 & 0.58 & 0.298 \\
\begin{tabular}[c]{@{}l@{}}Different \\ Source\end{tabular} & \begin{tabular}[c]{@{}l@{}}No \\ Eliminations\end{tabular} & 1.00 & 0.004 & 1.00 & 0.502 \\
\begin{tabular}[c]{@{}l@{}}Same \\ Source\end{tabular} & \begin{tabular}[c]{@{}l@{}}Made \\ Eliminations\end{tabular} & 0.046 & 0.036 & 0.25 & 0.152 \\
\begin{tabular}[c]{@{}l@{}}Same \\ Source\end{tabular} & \begin{tabular}[c]{@{}l@{}}No \\ Eliminations\end{tabular} & 0.000 & 0.000 & 0.15 & 0.078 \\ \bottomrule
\end{tabular}%
}
\end{table}

\replaced{T}{Generally, t}he original error rates reported for each black box study
follow Option 2: Inconclusives Correct. That is, inconclusives count in
the denominator of the error rate but not in the numerator (\citet[p. 7735-7736]{ulery2011}, \citet[p. 89]{monson2022accuracy}). In both
studies, this results in the smallest error rate for both same-source and
different-source pairs. Ignoring the inconclusives (Option 1) results in the same number of errors, but a smaller
denominator and a larger error rate. In the group of firearms
examiners who never excluded based on individual characteristics we find a 100\% false positive error rate and a 0\% false negative
rate when ignoring inconclusives, which is a natural consequence of grouping those examiners together. By definition, the `no individual eliminations' group never made exclusions based on individual characteristics, and so could never make a false exclusion on same-source comparisons or a true exclusion on different-source comparisons. By contrast, treating inconclusives as errors increases all error rates by at least seven fold,
and the error rate for non matches in the latent print study increases
by 210-fold. Scoring all inconclusive responses as erroneous is obviously extreme, but these massively inflated error rates speak to the scale of inconclusive responses in these studies. 

However, none of the error rates calculated above take the individual
response patterns of examiners and/or items into account. Consider two
items within an item bank: Item A for which every examiner who saw that
item reported it to be inconclusive, and Item B for which all but one
examiner came to the correct conclusive decision. Practical sense might
suggest that an inconclusive decision on Item B should be treated
differently from an inconclusive decisions on Item A.

Similarly, consider two examiners: Examiner X who reported inconclusive on
95\% of the items that they were shown and Examiner Y who reported
inconclusive on only 10\% of the items. It is possible for both of these
examiners to have made zero false positive or false negative errors, but
it is reasonable to believe that the inconclusives by Examiner Y are
more valid than those of 
Examiner X, as suggested by \citet{arkes-koehler2021}. 
Rather than applying the same scoring method to every inconclusive determination in every study indiscriminately, it might be more appropriate to account for these behaviors when computing error rates.

\section{Variance decomposition based on observed
responses}\label{sec-empirical-variance}

To formalize the idea introduced above, we begin by computing the
percentage of inconclusive responses for each examiner and item in both
studies. We then compute the variance of all of the examiner
inconclusive percentages, and the variance of the item inconclusive
percentages.

That is, for each study under consideration, compute $\hat{r}$, where

\begin{equation}
    \hat{r} = \frac{\hat{\sigma^2}_I}{\hat{\sigma^2}_I + \hat{\sigma^2_J}},
\end{equation}

\begin{equation}
    \hat{\sigma^2_I} = \text{Var}(\frac{1}{n_i} \sum_{j=1}^{n_i} \mathbb{I}\{X_{ij} = 1\}) \text{ and } \hat{\sigma^2_J} = \text{Var}(\frac{1}{n_j} \sum_{i=1}^{n_j} \mathbb{I}\{X_{ij} = 1\})
\end{equation}
where \(X_{ij} = 1\) if examiner
\(i\) reported an inconclusive on item \(j\). \(n_i\) is the number of
items seen by examiner \(i\) and \(n_j\) is the number of examiners who
responded to item \(j\). This results in a ratio, \(\hat r\), of the examiner
inconclusive variance ($\hat{\sigma^2}_I$) to the total inconclusive variance($\hat{\sigma^2}_I + \hat{\sigma^2}_J$). In order to
calculate adjusted error rates accounting for the patterns of
inconclusive in a particular study, we treat \(\hat r\)\ proportion of the total
inconclusives as erroneous.

While this begins to allow for the separation of examiner and item
variance, it is complicated by the fact that the \(n_i\)'s and \(n_j\)'s
vary wildly in the Monson study. Examiners responded to as few as 3 same
source items, and as many as 28 different source items. Particularly in small item-sets, this leads to massive uncertainty in the proportion of
inconclusive responses for an individual examiner. We return to this
issue in Section~\ref{sec-model-based-variance}.

\subsection{Results}\label{results}

The examiner variance, item variance, and ratio quantities are shown for the Ulery study in Table \ref{tbl-ulery-variance-decomp}. On same-source pairs, we attribute 6.5\% of the inconclusive responses to the examiners. On different-source pairs, we attribute 15.5\% of the inconclusive responses to the examiners. 

\begin{table}
\centering
\caption{Examiner variance, item variance, and ratio quantities for the Ulery (2011) study, separated by ground truth of items.}
\label{tbl-ulery-variance-decomp}

    \begin{tabular}{lrrr}
    \toprule
    Ground Truth & Examiner Var & Item Var & Ratio\\
    \midrule
    SS & 0.01 & 0.15 & 0.065\\
    DS & 0.02 & 0.11 & 0.155\\
    \bottomrule
    \end{tabular}
\end{table}

For the \citet{monson2022accuracy} study, we compute \(\hat r\) for each group of
examiners separately. Results are shown in Table~\ref{tbl-monson-variance-decomp}. On same-source pairs, we attribute 53\% of the inconclusive responses to the examiners in both groups, suggesting that
whether examiners exclude based on individual characteristics or not
does not impact their inconclusive rates on same-source pairs. On different-source pairs, we attribute 74\% of the inconclusive responses to the examiners in the group that made eliminations based on individual characteristics and 25\% of the inconclusive responses to the examiners in the group that did not exclude based on individual characteristics. 

Note that the proportions of inconclusives that are attributable to the examiners is much higher in the \citet{monson2022accuracy} study compared to the \citet{ulery2011}, suggesting that either latent print examiners may be more likely to agree with one another on inconclusive responses compared to firearms examiners; or there was much more variability in the item bank in the \citet{ulery2011} compared to the \citet{monson2022accuracy} study. 

\begin{table}
\centering
\caption{Examiner variance, item variance, and ratio quantities for the Monson (2023) study, separated by ground truth and examiner group.}
\label{tbl-monson-variance-decomp}

\begin{tabular}{llrrr}
\toprule
Ground Truth & Group & Examiner Var & Item Var & Ratio\\
\midrule
SS & Made Ind. Elims & 0.036 & 0.032 & 0.53\\
DS & Made Ind. Elims & 0.122 & 0.044 & 0.74\\
SS & No Ind. Elims & 0.042 & 0.037 & 0.53\\
DS & No Ind. Elims & 0.001 & 0.003 & 0.25\\
\bottomrule
\end{tabular}

\end{table}

\section{Model-Based Variance Decomposition}\label{sec-model-based-variance}

We now consider a formal statistical model to account for the issue of
examiners responding to different subsets of items. First, note that
\(X_{ij}\), which represents whether examiner $i$ reported a conclusive on item $j$, is a binomial random variable with a probability \(\pi_{ij}\) that
depends on both participant \(i\) and item \(j\).

\begin{equation}
    X_{ij} \sim \text{Binom}(\pi_{ij})
\end{equation}

Next, we assume that \(\pi_{ij}\) is a function of both participant
tendency to be conclusive (\(\theta_i\)) and item tendency to be judged conclusive (\(\zeta_j\)). These tendencies are latent variables, meaning they cannot be observed directly but are assumed to govern the observed responses. Any function that maps onto \([0,1]\) could be chosen, but we use the logistic function:

\begin{equation}
    \pi_{ij} =  \frac{1}{1+\exp(-(\theta_i + \zeta_j))}.
\end{equation}

Finally, we assume that \(\theta_i\) and \(\zeta_j\) are each drawn independently from some probability distributions. Rather than computing \(r\) with \(\hat{\sigma_p}\) and
\(\hat{\sigma_I}\), as in Section \ref{sec-empirical-variance}, in this section we use \(\text{Var}(\theta)\)
and \(\text{Var}(\zeta)\) directly, which do not depend on the item
sets given to each examiner. 

The model is therefore

\begin{equation}\protect\hypertarget{eq-rasch}{}{
P(\text{Examiner i is conclusive on Item j})) = P(X_{ij} = 1) =  \frac{1}{1+\exp(-(\theta_i + \zeta_j))}.
}\label{eq-rasch}\end{equation}

\begin{figure}
\centering \includegraphics[width=.7\linewidth]{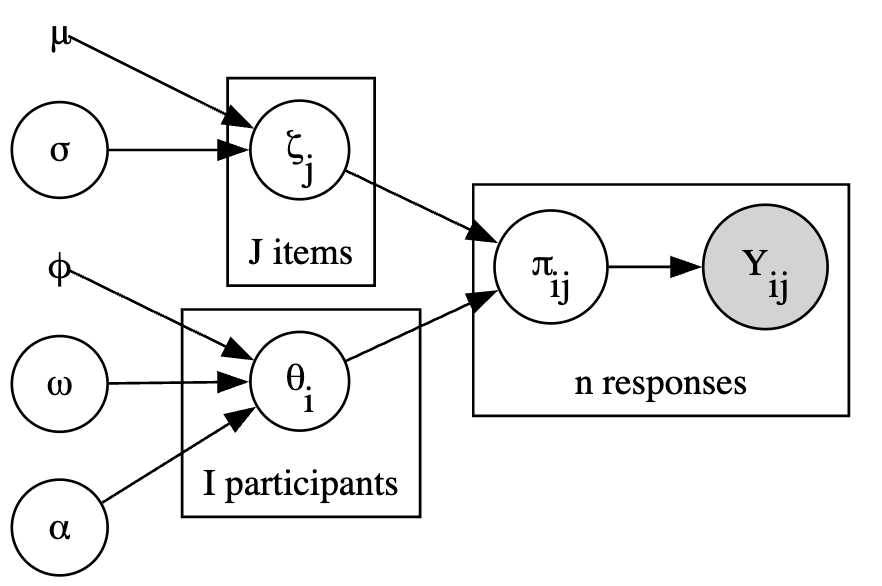}

\caption{\label{fig-graph-model} Graphical representation of the  model in Eq. \ref{eq-rasch}. Constants are represented as letters without shapes, random variables are represented as circles, and observed variables are represented with shaded circles. Rectangles represent variables that are repeated across items, participants, or responses.}

\end{figure}

This logistic probability model
is used often in standardized testing settings and falls under the
umbrella of Item Response Theory (IRT) (see, e.g., \citet{rasch1960studies, handbook2010} for further details). In that context, \(X_{ij}=1\) represents
participant \(i\) answering item \(j\) correctly, and \(\theta_i\) and
\(\zeta_j\) represent participant proficiency and item easiness,
respectively. For further details and examples of IRT applied in
forensic science, see \citet{luby2018proficiency, luby2023method}.

The formula in Equation~\ref{eq-rasch} suffers from an identification
problem, since a constant could be added to each \(\theta_i\) and subtracted from each
\(\zeta_j\) without changing the resulting probabilities. To address
this issue, we \deleted{will} center the mean of the item tendencies at zero.

We also assign

\begin{equation}
{\theta}_i \sim \text{Skew-Normal}(0, \omega, \alpha) \text{ and } {\zeta}_j \sim N(0, \sigma_\zeta),
\end{equation}

\noindent which allows the shape of the examiner tendencies to be skewed \citep{azzalini1985skewnormal}. \added{This is desireable, for example, in settings where most examiners report few inconclusives, but a few examiners report many inconclusives.} The mean of the examiner tendencies is then
\begin{equation}
    \mu_\theta = \omega \sqrt{\frac{2}{\pi}} \frac{\alpha}{\sqrt{1+\alpha^2}, }
\end{equation}

\noindent with the variance is given by

\begin{equation}
    \sigma^2_\theta = \omega^2(1 - \frac{2\alpha^2}{\pi(1+\alpha^2)}).
\end{equation}

\begin{figure}

{\centering \includegraphics{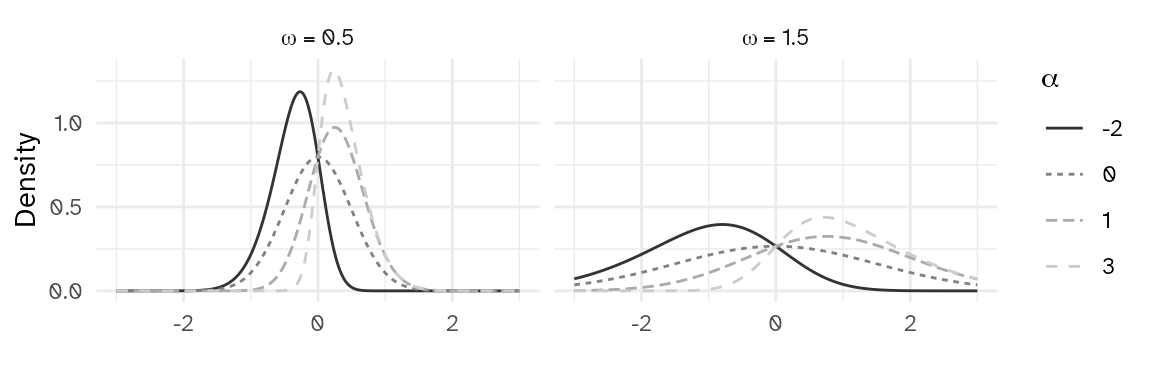}

}

\caption{\label{fig-skew-normal-dists}Examples of various skew-normal
distributions.}

\end{figure}

Some example skew normal distributions are shown in
Figure~\ref{fig-skew-normal-dists}. When \(\alpha = 0\) and $\omega=1$, we obtain the
standard normal distribution. Otherwise, \(\alpha\) controls the
direction and amount of skew of the distribution and $\omega$ controls the spread.

We also assign the following hyperpriors:

\begin{equation}
\sigma_\zeta  \sim \text{Half-}T_3, \omega \sim \text{Half-}T_3, \text{ and } \alpha \sim T_3,
\end{equation}

\noindent where Half-T$_3$ refers to a \added{zero-}truncated Student's $t$ distribution with three degrees of freedom. This choice guarantees non-negative variance estimates for the items and examiners. The $t$ distribution with three degrees of freedom is symmetric and bell shaped, with heavier tails than the normal distribution. \added{Using three degrees of freedom results in a finite mean and variance of the prior distribution, while using one or two degrees of freedom would not.}

\begin{figure}

{\centering \includegraphics{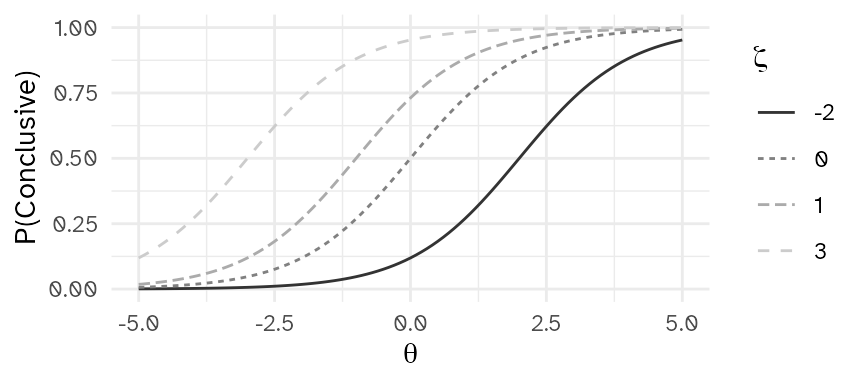}

}

\caption{\label{fig-example-icc}Illustration of the model for
P(Conclusive) for four hypothetical items ($\zeta = -2, 0, 1, 3$), for  plausible values of $\theta$.}

\end{figure}

Figure~\ref{fig-example-icc} displays P(Conclusive) from the model above
for four different hypothetical items. Items are more likely to be rated
as inconclusive if they have a lower \(\zeta\) estimate, and examiners
are more likely to report inconclusives if they have a lower \(\theta\)
estimate. When \(\zeta= \theta\), the probability of a conclusive
response is $0.5$.

\subsection{Results}\label{sec-results}

After applying the model in
Section~\ref{sec-model-based-variance} to each dataset\footnote{Some known/questioned pairs were assigned in multiple rounds to a single examiner in \citet{monson2022accuracy}. In these cases, we use the first response when fitting the model.} in
Section~\ref{sec-two-black-box-studies}, we obtain parameter estimates
for each participant (\(\theta_i\)) and item (\(\zeta_j\)), along with
hyperparameters for the mean and standard errors. Each model was fit using stan \citep{stan} within R \citep{rstan, Rsoftware} with 4 chains and 5000 iterations per chain, discarding the first 2500 iterations. 

\subsubsection{\replaced{Model-Based Ratio Quantities}{Posterior
standard errors}}\label{sec-posterior-standard-errors}

\deleted{A simple way to assess whether inconclusive responses in each study are
overall more attributable to the examiners or to the items is to compare
their posterior standard errors. If \(\sigma_\theta < \sigma_\zeta\),
the items as a whole have more variability in their tendency to be
conclusive and we may expect more inconclusive responses due to the
items than to the examiners. On the other hand, if
\(\sigma_\theta > \sigma_\zeta\), the examiners have more variability
than the items, and we may expect more inconclusive responses due to the
examiners than to the items.}

\deleted{In the latent prints study, the mean estimate of the posterior standard deviation (with 95\%
posterior credible intervals) is \(\sigma_\theta = 1.9 [1.46,2.47]\) on
same source items and \(\sigma_\theta = 4.06 [3.04,5.48]\) on different source items. On
same-source items, \(\sigma_\zeta = 26.82 [22,32.83]\), and on
different-source items \(\sigma_\zeta = 16.36 [12.47,21.59]\),
suggesting that the items play a larger role in inconclusive responses
overall.}

\deleted{The bullets study resulted in \(\sigma_\theta = 0.58 [0.39,0.84]\) and
\(\sigma_\zeta = 3.76 [2.02,6.8]\) on same source items, suggesting that
the examiners may play a smaller role in inconclusive responses than the
items. On the different source items,
\(\sigma_\zeta = 0.59 [0.36,0.93]\). We have two \(\sigma_\theta\)
values for different source items, one for the group who did exclude based on individual
characteristics: \(\sigma_\theta = 0.99 [0.96,1]\); and one for the
group who did not exclude based on individual characteristics:
\(\sigma_\theta = 0.51 [0.38,0.73]\). For the group who did exclude based on individual characteristics, examiners may play a larger role in inconclusive responses than the items; and in the group that did not exclude based on individual characteristics, the examiners and the items are play a similar role in inconclusive responses.}

\deleted{Since the posterior distributions account for the fact that different
examiners responded to different subsets of items, the posterior
standard errors are a more reliable measure of the overall variability
across examiners and items than the observed standard errors reported in Section \ref{sec-empirical-variance}.}

\added{The model-based examiner variance, item variance, and ratio quantities for the \citet{ulery2011} study are shown in Table~\ref{tbl-ulery-variance-decomp-model}. While the examiner and item variances are much larger than the observed variances in Table~\ref{tbl-ulery-variance-decomp}, the ratio quantity for same-source pairs is very similar. The ratio quantity for different-source pairs is about 4 percentage points higher using the model-based estimation.}

\added{The same quantities for the \citet{monson2022accuracy} are shown in Table~\ref{tbl-monson-variance-decomp-model}. The model-based ratio quantities are smaller than the observed ratio quantities on same-source items, and larger than the observed ratio on different-source items. The difference is especially pronounced for the group of examiners who did not make eliminations based on individual characteristics.}

\begin{table}

\caption{Model-based examiner variance, item variance, and ratio quantities for the Ulery (2011) study, separated by ground truth. The observed ratio quantity is also included for comparison.}
\label{tbl-ulery-variance-decomp-model}
\begin{tabular}{lrrrr}
\toprule
Ground Truth & Obs Ratio & Item Var (Model) & Examiner Var (Model) & Ratio (Model) \\
\midrule
SS & 0.065 & 26.776 & 1.908 & 0.067\\
DS & 0.155 & 16.282 & 4.066 & 0.200\\
\bottomrule
\end{tabular}

\end{table}

\begin{table}

\caption{Model-based examiner variance, item variance, and ratio quantities for the Monson (2023) study, separated by ground truth and examiner group. The observed ratio quantity is also included for comparison.}
\label{tbl-monson-variance-decomp-model}
\begin{tabular}{llrrrr}
\toprule
Ground \tabularnewline Truth & Group & Obs Ratio & Item Var (Model) & Examiner Var (Model) & Ratio (Model)\\
\midrule
SS & Made Ind. Elims & 0.531 & 2.317 & 1.343 & 0.367\\
DS & Made Ind. Elims & 0.736 & 1.367 & 9.054 & 0.869 \\
SS & No Ind. Elims & 0.530 & 2.317 & 2.812 & 0.548 \\
DS & No Ind. Elims & 0.253 & 1.367 & 261 & 0.995 \\
\bottomrule
\end{tabular}

\end{table}

\subsubsection{Constructing uncertainty intervals for observed
quantities}\label{constructing-uncertainty-intervals-for-observed-quantities}

Upon first reading, the model-based ratio quantities might appear to contradict the empirical ratios \added{reported in Tables \ref{tbl-ulery-variance-decomp} and \ref{tbl-monson-variance-decomp}}. However, the discrepancy is due to different examiners responding to different item-sets. The model-based estimates of examiner tendency to be conclusive ($\theta$) and item tendency to be rated conclusive ($\zeta$) account for these different item-sets, and are therefore measures of tendency as if every examiner responded to every item. 

The model-based estimates of $\theta$ and $\zeta$, can be used to construct uncertainty intervals for the observed $\hat{r}$ quantity computed in Section \ref{sec-empirical-variance}. The following intervals were calculated using a posterior predictive check: for each draw of $\theta$ and $\zeta$ from
the posterior distribution, we predict an observed response for each
examiner \(\times\) item pair that was assigned in the original study.
We therefore simulate \(n\) possible outcomes of each study, where \(n\)
is the number of posterior draws. For each simulated study, we can
compute \(\hat{r}\) based on the observed \(\hat\sigma_i\) and
\(\hat\sigma_j\). By looking at the 95\% intervals of \(\hat{r}\), we find a
95\% uncertainty interval for the \(\hat{r}\) that was observed.

For the bullets data, we obtain the intervals in Table \ref{tbl-monson-pred-vs-observed-ratio}. All 95\% posterior intervals contain the observed ratio quantity, and the intervals are larger for the group that did not make eliminations based on individual characteristics, indicating more uncertainty for the smaller group of examiners. 

\begin{table}
\centering
\caption{Predicted and observed ratio estimates for the \citet{monson2022accuracy} study, with 95\% posterior intervals, separated by ground truth and examiner group. }
\label{tbl-monson-pred-vs-observed-ratio}
\begin{tabular}{llrrrr}
\toprule
Ground truth & Group & Predicted Ratio & Lower & Upper & Observed Ratio\\
\midrule
DS & Made Ind. Elims & 0.754 & 0.713 & 0.793 & 0.736 \\
DS & No Ind. Elims & 0.365 & 0.070 & 0.652 & 0.253 \\
SS & Made Ind. Elims & 0.501 & 0.399 & 0.611 & 0.531 \\
SS & No Ind. Elims & 0.513 & 0.347 & 0.688 & 0.530 \\
\bottomrule
\end{tabular}
\end{table}

For the latent print data, we obtain the intervals in Table \ref{tbl-ulery-posterior-pred}. The predicted ratios are indistinguishable from the observed ratios, and the uncertainty intervals are substantially narrower than those for the \citet{monson2022accuracy} study. 

\begin{table}
\centering
\caption{Predicted and observed ratio estimates, with 95\% intervals for the Ulery (2011) study.}
\label{tbl-ulery-posterior-pred}
\begin{tabular}{llrrrr}
\toprule
Ground Truth & Predicted Ratio & Lower & Upper & Observed Ratio \\
\midrule
DS & 0.155 & 0.133 & 0.178 & 0.155 \\
SS & 0.065 & 0.057 & 0.074 & 0.064 \\
\bottomrule
\end{tabular}
\end{table}

Since each of the posterior prediction intervals for $r$ contain the observed ratio from the study, the discrepancy between the observed ratio quantity and the model-based ratio quantity (as displayed in Tables \ref{tbl-ulery-variance-decomp-model} and \ref{tbl-monson-variance-decomp-model}) reflects the differing item-sets given to each examiner. 

\subsubsection{Ratio-adjusted \replaced{failure}{error}
rates}\label{ratio-adjusted-error-rates}

Finally, we can use the model-based ratio estimates of $r$ to adjust the error rates within each study. \added{We call these the "failure rates" to reflect that they do not represent the rates of false positive or false negative errors. Rather, they represent the overall rates of failing to come to the correct decision on same source or different source comparisons.} These are shown in Table \ref{tab-ulery-adj-error-rates-model} for the \citet{ulery2011} study; and Table \ref{tab-monson-adj-error-rates-model} for the \citet{monson2022accuracy} study. \added{In both studies, the model-adjusted failure rates track closer to the reported error rates (Inconc Correct column) for ground truth matches, reflecting the smaller values of the model-based ratio quantity. On different source pairs, however, there is a substantial increase compared to the reported error rates. For the latent prints data, the failure rate on different source pairs of over 4\%, 40 times larger than the reported false positive error rate. For the bullets data, the failure rate on different source pairs of 50.7\% for the group that made eliminations based on individual characteristics, and 99.5\% for the group that made no eliminations based on individual characteristics.  This suggests that, on different source pairs especially, there is significant examiner variability in reporting inconclusive responses.}

\added{The failure rates based on the model-based quantities are very similar to the failure rates based on the observed quantities in the latent print study. In the bullets study, however, these differences are much larger, particularly on different-source comparisons. Accounting for the item-sets assigned to each participant therefore reveals additional variability in the inconclusive patterns.}

\deleted{It is important to note that the model-adjusted error rates reported here are \textit{not} equivalent to false positive or false negative error rates. Rather, they are overall error rates on different source and same source comparisons that include false positive and negative errors, as well as inconclusives that have been attributed to the examiners.}

\begin{table}
\centering
\caption{Error and failure rate comparison for the \citet{ulery2011} data. `No value' determinations have been excluded from these calculations, following the original study.}
\label{tab-ulery-adj-error-rates-model}
\begin{tabularx}{\textwidth}{lXXXX}
\toprule
Ground Truth &  Inconc Correct & Inc Incorrect & Obs Ratio Failure Rate & Model Adjusted Failure Rate\\
\midrule
SS & 0.075 & 0.548 & 0.105 & 0.106 \\
DS & 0.001 & 0.208 & 0.033 & 0.042 \\
\bottomrule
\end{tabularx}
\end{table}

\begin{table}
\centering
\caption{Error and failure rate comparison for the \citet{monson2022accuracy} data. `Unsuitable' determinations have been excluded from these calculations, following the original study. }
\label{tab-monson-adj-error-rates-model}
\begin{tabularx}{\textwidth}{lrXXXX}
\toprule
Ground Truth & Group & Inc Correct Error Rate & Inc Incorrect Error Rate & Obs Ratio Failure Rate & Model Ratio Failure Rate\\
\midrule
SS & Made Ind. Elims & 0.036 & 0.255 & 0.152 & 0.116\\
SS & No Ind. Elims & 0.000 & 0.153 & 0.081 & 0.084\\
DS & Made Ind. Elims & 0.008 & 0.583 & 0.431 & 0.507\\
DS & No Ind. Elims & 0.004 & 1.000 & 0.256 & 0.995\\
\bottomrule
\end{tabularx}
\end{table}

\section{Discussion}\label{sec-discussion}

In feature comparison disciplines, it is common to encounter
inconclusive results. However, the likelihood of
inconclusive results varies across comparisons, examiners, and studies.
How inconclusives should be treated within the `error rate' framework
has been argued extensively. \deleted[]{While it may be most appropriate to report inconclusive rates alongside error rates for same and different source items, it is often desirable to summarize overall performance within a study with a single number. }  In this
paper, we have presented one framework for \replaced{computing "failure rates"}{adjusting reported error
rates} based on inconclusive responses that depends on the overall
pattern of inconclusive responses for each particular study. By
separating examiner tendencies from item tendencies, this framework
provides a principled way of determining what proportion of inconclusive results are due to the item versus examiner tendencies.

Framing the issue in terms of the ratio of examiner variability to total variability also provides a metric for comparing `black box' study results within and across disciplines. We've shown that these ratios vary substantially between \added{two} well-known latent print and firearm studies. One could imagine future rigorously-designed black box studies in these disciplines resulting in different ratio quantities depending on participant samples or item bank construction. Studies with low ratio quantities indicate more variable item banks, while studies with high ratio quantities indicate more variability across examiners. Furthermore, studies with model-based ratios close to observed ratios (such as the \citet{ulery2011} study) indicate individual item-sets are both large and representative enough to estimate individual tendencies with observed proportions. Studies with model-based ratios substantially different than observed ratios (such as the \citet{monson2022accuracy} study) may indicate studies that have not assigned item-sets to individuals that are both large and representative enough to estimate individual tendencies with observed proportions. Critically, estimating these ratios requires data to be available at the response level (i.e. how each participant responded to each item), which is not the case for the vast majority of existing black box studies (see Appendix \ref{appendix-other-black-box-studies-in-feature-comparison-disciplines}). Future studies should follow the lead of \citet{ulery2011} and \citet{monson2022accuracy} (among others) by making response-level data available. 

\subsection{Latent Prints Examination}\label{feature-comparison-disciplines}

When computing quantities for the \citet{ulery2011} study, the model-based ratio and ratio based on observed rates were remarkably similar. This could be due to the large number of items that were assigned to each participant (roughly 100) and each participant being assigned a similar range of items. This demonstrates the importance of assigning large and varied item sets in error rate studies: it decreases the amount of uncertainty in the resulting estimates. 

However, the resulting ratio quantity does suggest substantial examiner variability in making inconclusive decisions, particularly on ground truth non-matches. Since the reported error rates are so small, incorporating this variability into the error rates results in \replaced{a failure}{an error} rate on non-matches that is  43 times larger than the reported error rate. 

\subsection{Firearms Examination}\label{firearms-examination}

In the \citet{monson2022accuracy} study, the model-based ratios were substantially different than the ratios based on observed rates of inconclusives. This suggests that the set of items seen by each examiner was not large and/or representative enough to accurately estimate their tendency to be inconclusive with the observed quantities. This could be due to the smaller item set size (generally $15$ or $30$) or major differences in the number of low-quality items seen.  

Additionally, in
firearms examination, some laboratories (including the FBI) do not allow
exclusions based on individual characteristics without access to the
physical firearm. In black box studies involving only items that
come from the same brand of firearm and ammunition, such participants will
never report \emph{correct} exclusions. \citet{scurich2023commentary} notes
that roughly one-third of all cartridge case eliminations in the \citet{monson2022accuracy} study were reported to be based on class characteristics,
despite the study claiming that all pairs matched on class
characteristics. While this can likely be attributed to different
examiners using different definitions of `class characteristic', this
variability does \replaced{raise questions about how applicable the reported error rates are to a casework setting}{cast doubt upon whether the single error rate can
truly capture the performance of an entire discipline}.

In a variety of recent black box studies in firearms, the data has
supported that examiners tend to make more inconclusive decisions on
ground truth non-matches relative to ground truth matches \citep{baldwin2014study, keisler2018isolated, monson2022accuracy, best2020assessment}. This
could be interpreted as they are more confident in making identification
decisions, but more conservative when making exclusions. According to
\citet{sinha-gutierrez2023}, prosecutors may use inconclusive results as
evidence of guilt, \added{which may impinge on an examiner's decision}. Our analysis has shown that accounting for
inconclusive results attributable to examiners 
leads to \replaced{failure}{an error} rates that are substantially higher than reported error
rates.

\subsection{\replaced{Challenges in Interpreting Study Results}{Legal Considerations}} \label{legal-conclusions}

The territory we have addressed is hotly disputed in the literature \citep{dror2019cannot, hofmann2020treatment, dror2020mis, dror2022, dorfman2022inconclusives, arkes-koehler2021, arkes-koehler2022, scurich2022, swofford2024inconclusive}. 
Lower error rates bolster the credibility of the forensic technique tested, while higher error rates diminish that credibility.

Forensic analysts are human beings. As such, mistakes are made. \textit{Any} research study on subjective human decision-making with a mistake rate of one in a thousand is implausibly low, even for very simple tasks. \added{For example, the human error rate on image-based CAPTCHA tasks has been estimated at 7\%} \citep{bursztein2010good}. \added{As a higher-stakes example, the error rate for diagnostic radiology is often estimated to be between 2-4\% } \citep{graber2013incidence}. The \citet{ulery2011} report of an error rate of one in a thousand for different source items strikes us as so low as to strain credibility. The same source error rate of 7.5\%, 75 times greater, does not strike us in the same way. Similarly, with respect to the \citet{monson2022accuracy} study, the reported error rates of 0.008 and 0.004 for different sources, again strike us as implausibly low, while the rate of 0.036 for same source items (among those who excluded at least one item), feels more plausible. Again, this is not a comment on the forensic tasks, but a comment on expected human fallibility.

It's unclear whether the estimated error rates from black box studies are applicable to casework settings. 
Many local, state, and federal forensic laboratories are bureaucratically part of, and answerable to, the police and prosecution. 
Even if there is no explicit pressure as a result, the hierarchical relationships can unintentionally influence outcomes in favor of convictions. 
As a consequence, some examiners may be more aggressive, determining "same source" in situations that reflect some uncertainty. 
Both \citet{ulery2011} and \citet{monson2022accuracy} were designed and conducted in the wake of the acceptance of DNA analysis as a scientifically-based forensic tool. 
In both studies, it is reasonable to suppose that participants understood that these experiments could bolster their discipline if the results came out "right", that is, with low false positive error rates. This could manifest in more cautious decisions, and higher inconclusive rates than would be expected in casework \citep{orne1962}. 
     
Alternatively, one could argue that an experiment involves lower stakes for a forensic examiner than routine casework, since only the latter has consequences for the life and liberty of the defendant, and the examiner's own professional standing. This argument suggests that examiners might be more aggressive in an experiment than in casework, leading to higher rates of false positives than routine practice. Considering the rarity of blind testing in pattern evidence disciplines \citep{mejia2020implementing}, it is not obvious how behavior might change as a result of being in an experiment.

In court, error rates from black box studies are used to justify the inclusion of forensic testimony. However, these error rates may be calculated in a variety of different ways, and how inconclusive results are included or excluded from these calculations has a massive impact on the resulting rate. The proposed "failure rates" on same and different source comparisons can help ensure that black box study results are appropriately weighed in legal proceedings. \deleted{Thus
     the best we can do for the court is to point to this likely
     source of bias and uncertainty, and leave it to the fact-finder (judge or jury)
     to decide how much credence to give to these test results.}

\deleted{Presenting the inconclusive rates alongside the error rates for same and different source items would be the most precise way to include such information. However, multiple quantities describing a single study introduces complexity and so it is also desirable to summarize each study with a single quantity.  The Hawthorne Effect literature does not advise us on how
     behavior might be different in an experiment that it would be in
     routine work. The social context literature does allow us to
     predict the direction of the effect, but not its magnitude.}

\subsection{Conclusion}\label{conclusion}

Inconclusive results are to be expected in feature comparison
disciplines, and play a crucial role in minimizing false
positive decisions. However, when misused, they can also deprive defendants of exculpatory evidence. They do not fit neatly into the ``error
rate'' framework that has become the practice in the United States for
evaluating the performance of feature comparison disciplines, but
incorporating inconclusive decisions can massively change the error
rates depending on whether they count in the numerator, denominator, or
neither. Furthermore, inconclusive results are not equally likely on
every comparison or among every participant, and so treating all
inconclusive results the same is an oversimplification.

In this paper, we argue that the overall pattern of inconclusive
decisions in a particular study can \replaced{be used to compute `failure rates'}{inform the inconclusive-adjusted error rates.} 
Using a latent variable based approach, some proportion of inconclusive responses can be attributed to participants, and can be
considered erroneous. The inconclusive proportion that can be attributed
to the designers' choice of items are not considered erroneous to the participants. Importantly, this approach results in \replaced{failure}{error} rates that are adjusted for each study individually. Studies with more difficult items that most participants agree should be \added{reported as} inconclusive will have smaller adjusted \replaced{failure}{error} rates than studies with lots of disagreement among participants. We have contrasted three different ways of thinking about errors in
     the two studies we address; the reported error rate, the
     empirical-variance-corrected failure rate and the model-corrected
     failure rate.  If we were forced to choose among them, we
     think that the model-based failure rate does a better job of
     protecting against the effects of study design,
     both in the choice of study items and in the assignment of study items to individual participants. \added{This approach, which considers the contributions of both participants and items to inconclusive responses, offers a new viewpoint for understanding the issue.}

\deleted{Inconclusive tendencies are another aspect of performance
that should be studied alongside error rates and regularly measured. In most existing black box studies, it is impossible to distinguish inconclusive decisions where the participant may have been considering the wrong decision (and therefore the inconclusive prevented an error) from an inconclusive decision . A notable exception is \citet{mattijssen2020validity}, which forced participants to first make a same or different source conclusion, and then asked whether they would report an inconclusive if they encountered that item in casework. 
A further issue with measuring inconclusive tendency is that reporting standards for inconclusive results vary across
forensic laboratories. In order to minimize variability, standardization across laboratories is desirable.}

These results illustrate the profound impact that study
design can have on the number of inconclusive, erroneous, and correct
responses. The item bank and how questions are assigned have
major impacts on observed responses. Crucially, determining the impact of these aspects of study design requires individual-level responses. It would be helpful if all black box studies routinely made available the determinations made by each participant on each item, as did \citet{ulery2011} and \citet{monson2022accuracy}. 

\section*{Funding}

Luby was supported by the Center for Statistics and Applications in Forensic Evidence (CSAFE) through Cooperative Agreements 70NANB15H176 and 70NANB20H019 between National Institute of Standards and Technology and Iowa State University, which includes activities carried out at Carnegie Mellon University, Duke University, University of California Irvine, University of Virginia, West Virginia University, University of Pennsylvania, Swarthmore College and University of Nebraska, Lincoln.

\section*{Acknowledgments}

We are grateful to Richard Gutierrez for many helpful comments on this manuscript. 

\bibliography{lpr-inconclusives}

\section*{Appendix A: Study
Comparison}\label{appendix-a-study-comparison}

\begin{longtable}[]{@{}
  >{\raggedright\arraybackslash}p{(\columnwidth - 4\tabcolsep) * \real{0.2000}}
  >{\raggedright\arraybackslash}p{(\columnwidth - 4\tabcolsep) * \real{0.4000}}
  >{\raggedright\arraybackslash}p{(\columnwidth - 4\tabcolsep) * \real{0.4000}}@{}}
\toprule\noalign{}
\begin{minipage}[b]{\linewidth}\raggedright
\end{minipage} & \begin{minipage}[b]{\linewidth}\raggedright
Ulery et al. (2011)
\end{minipage} & \begin{minipage}[b]{\linewidth}\raggedright
Monson et al. (2022)
\end{minipage} \\
\midrule\noalign{}
\endhead
\bottomrule\noalign{}
\endlastfoot
\emph{Item Bank Size} & 744 & 228 \\
\emph{Ground Truth Matches} & 70\% & 17\% \\
\emph{Item Selection} & \emph{Image pairs were selected to be
challenging: Mated pairs were randomly selected from the multiple
latents and exemplars available for each finger position; nonmated pairs
were based on difficult comparisons resulting from searches of IAFIS} &
3 firearm manufacturers and a single brand of ammunition \emph{\ldots.
were selected for their propensity to produce challenging and ambiguous
test specimens, creating difficult comparisons for examiners.} \\
\emph{Comparison to casework} & \emph{Participants were surveyed, and a
large majority of the respondents agreed that the data were
representative of casework} & \emph{evidentiary specimens may generally
be assumed to be less challenging than those used in this study.} \\
\emph{Items per Examiner} & 98-110 (Mode of 100)  & 15
(59 participants)

30 (113 participants)

45 (1 participant) \\
\emph{Number of Participants} & 169 & 173 \\
\emph{Sampling Method} & Mostly voluntary, some encouraged or required
to participate by their employers. & Voluntary: calls for participation
made through professional organizations and email listservs. \\
\emph{Examiner Population} & All latent print examiners & United States,
non-FBI examiners only \\
\emph{Employment} & 48\% employed by U.S. federal agencies

23\% US state agencies

21\% US local agencies

7\% private organization

1\% non-US organizations & 
2 (2.5\%) US Federal 

46 (58.2\%) US State

31 (39.2\%) US Local 

\emph{Note:} total does not equal reported sample size (see \citet{bajic2020report})

\\
\emph{Responses} & \textbf{Individualization}: \emph{The two
fingerprints originated from the same finger.}

\textbf{Inconclusive}: \emph{Neither individualization nor exclusion is
possible.} (see below for reasons)

\textbf{Exclusion:} \emph{The two fingerprints did not come from the
same finger.}

\textbf{No Value:} latent print image was not ``of value for
individualization'' or ``of value for exclusion only'' &
\textbf{Identification}: \emph{Agreement of a combination of individual
characteristics and all discernible class characteristics where the
extent of the agreement exceeds that which can occur in the comparison
of toolmarks made by different tools and is consistent with the
agreement demonstrated by toolmarks known to have been produced by the
same tool.}

\textbf{Inconclusive} (see below)

\textbf{Elimination}: \emph{The significant disagreement of discernible
class characteristics and/or individual characteristics.}

\textbf{Unsuitable} for Examination \\
\emph{Inconclusive Types} & \textbf{Close}: \emph{The correspondence of
features is supportive of the conclusion that the two impressions
originated from the same source, but not to the extent sufficient for
individualization.}

\textbf{Insufficient}: \emph{Potentially corresponding areas are
present, but there is insufficient information present.}~Examiners were
told to select this reason if the reference print was not of value.

\textbf{No} \textbf{Overlap}: \emph{No overlapping area between the
latent and reference prints} & \textbf{Inc-A}: \emph{Some agreement of
individual characteristics and all discernible class characteristics,
but insufficient for an identification.}

\textbf{Inc-B:} \emph{Agreement of all discernible class characteristics
without agreement or disagreement of individual characteristics due to
an absence, insufficiency, or lack of reproducibility.}

\textbf{Inc-C}: \emph{Agreement of all discernible class characteristics
and disagreement of individual characteristics, but insufficient for an
elimination.} \\
\end{longtable}

\pagebreak 

\section*{Appendix B: Other Black Box Studies in Feature Comparison
Disciplines}\label{appendix-other-black-box-studies-in-feature-comparison-disciplines}

\FloatBarrier

\begin{table}[h]
\begin{tabular}{lllllll}
\toprule
Study & Discipline & Examiners & Items & SS & DS & \parbox{1.5cm}{\centering Individual\\Responses\\Published} \\
\midrule
\citet{ulery2011} & Latent Fingerprints & 169 & 744 & 70\% & 30\% & Yes \\
\citet{monson2022accuracy} & Bullets & 173 & 228 & 17\% & 83\% & Yes  \\
\citet{ELDRIDGE2021110457} & Palm Prints & 226 & 526 & 76\% & 24\% & Yes \\
%Monson & Bloodstain Pattern Analysis & 75 & 192 & NA & NA & No  \\
\citet{fadul2013empirical} & Firearms/toolmarks & 165 & 10 & 8\tablefootnote{Rather than each item consisting of one unknown sample to one or more known samples, this study asked examiners to select one of eight known sources for the entire set of 10 unknown sources.} & 2 & No \\
\citet{richetelli2020forensic} & Footwear & 70 & 12 & 5 & 7 & No  \\
\citet{chapnick2021results} & VCM & 76 & 40 & 17 & 23 & No \\
\citet{baldwin2023study} & Firearms & 218 & 15 & 5 & 10 & No \\
\bottomrule
\end{tabular}
\end{table}

\end{document}